# Ferromagnetic Spin Glass State and Anomalous Hall Effect in Topological Semimetal Candidate $Mn_2Sb_2Te_5$

M.M. Sharma[1], Ankush Saxena[2], S.M. Huang[2], Santosh Karki Chhetri[1], Jin Hu[1,3#], V.P.S. Awana[4*]

*[1] Department of Physics, University of Arkansas, Fayetteville, AR 72701, USA*
*[2] National Sun Yat-sen University, Kaohsiung, Taiwan - 804*
*[3] Institute for Nanoscience and Engineering, MonArk NSF Quantum Foundry, and Smart Ferroic Materials Center, University of Arkansas, Fayetteville, Arkansas 72701, USA*
*[4] CSIR- National Physical Laboratory, New Delhi 110012, India.*

**Abstract:**

Materials that intrinsically possess both magnetism and topological states represent a key frontier of quantum materials research. Recently, $Mn_2(Bi/Sb)_2Te_5$ has emerged as a promising candidate for hosting topological surface states coupled with intrinsic magnetic order, making it a potential magnetic Weyl semimetal. In this study, we investigate the magnetic and transport properties of $Mn_2Sb_2Te_5$ single crystals. The magnetization measurements reveal a spin glass state with field-induced ferromagnetism. Although heat capacity measurement indicates the absence of long-range order, the intrinsic magnetization in $Mn_2Sb_2Te_5$ significantly affects its electrical properties, as demonstrated by the anomalous Hall effect. This work provides valuable insights into the magnetism and the electronic properties of $Mn_2Sb_2Te_5$, establishing $Mn_2(Bi/Sb)_2Te_5$ system as a compelling platform for exploring the interplay between magnetism and non-trivial band topology, enabling emergent quantum phases and novel transport responses not accessible in non-magnetic systems.

[#]jinhu@uark.edu

[*]awana@nplindia.org

**Introduction**

Topological materials have attracted extensive attention recently due to their remarkable properties. A defining characteristic of these materials is the presence of conducting surface/edge topological states in topological insulators, while the nature of the bulk energy gap plays a critical role in shaping their exotic behaviours [1–4]. Magnetic topological insulators are a distinct class that exhibit a magnetically ordered bulk phase alongside the topological surface states [3,5,6]. The interplay between non-trivial band topology and bulk magnetic ordering in magnetic topological insulators leads to unique transport phenomena such as the quantum anomalous Hall effect [6–8], axion insulator state [9–11], and topological magneto-electric effect [12,13]. Beyond their fundamental interest, magnetic topological materials are also of considerable technological importance [3,5]. The coupling between magnetic order and topological electronic states enables control of Berry curvature and spin-polarized transport through external magnetic fields or magnetic engineering [3,6]. Such tunability can lead to dissipationless edge conduction, robust anomalous Hall responses, and magnetically switchable topological phases [6-8]. These properties make magnetic topological materials promising candidates for next-generation spintronic devices, topological memory elements, and low-power quantum electronic technologies [3,5]. Recent efforts have been focused on inducing bulk magnetic ordering in topological materials, including the fabrication of heterostructures consisting of magnetic and topological insulator layers [14–16] and doping topological insulators with magnetic atoms [17–19]. These methods, however, present challenges such as experimental complexity and defects. The challenges can be overcome by developing intrinsic materials with bulk magnetic ordering and non-trivial band topology. Topological insulators such as $Bi_2Se_3$ and $Sb_2Te_3$ feature layered structures separated by van der Waals gaps, allowing for the insertion of additional atomic layers to create natural heterostructures, offering a versatile platform for tuning the quantum properties of topological insulators [20–23].

In this context, the Mn(Bi/Sb)$_{2n}$Te$_{3n+1}$ (n=1,2,3…) family has been extensively studied. These materials are formed from MnTe layers embedded within multiple (Bi/Sb)$_2$Te$_3$ unit cells [24–28]. The MnTe layers induce antiferromagnetic (AFM) or ferromagnetic (FM) ordering in the system, depending on the nature of the coupling of Mn spins [29–31]. The first member in this series, MnBi$_2$Te$_4$, is an AFM magnetic topological insulator that hosts a magnetic field-driven Weyl state [26] and exhibit quantum anomalous Hall effect [8]. Theoretical studies also suggest that MnBi$_2$Te$_4$ can host a topological axion state [11,32]. The Sb-based compound, MnSb$_2$Te$_4$, displays field-induced FM ordering with spin glass features [33,34] and can be tuned to a Weyl semimetal under pressure [35].

So far, most studies have focused on compounds with a single MnTe layer in (Bi/Sb)$_2$Te$_3$ cells, i.e., MnBi$_2$Te$_4$ [8,32], while experimental studies on these with multiple MnTe layers remain less explored. Theoretical studies suggest that Mn$_2$Bi$_2$Te$_5$ with two MnTe layers in (Bi/Sb)$_2$Te$_3$ cells could provide an ideal platform for realizing an axion insulator phase [36,37]. Recent study reports the successful synthesis of Mn$_2$Bi$_2$Te$_5$ single crystals with an ABC-type stacking of nine atomic layers (Te-Bi-Te-Mn-Te-Mn-Te-Bi-Te), where two Mn atoms have opposite spin orientations that give rise to an AFM ordering [38]. The Sb-based compound, Mn$_2$Sb$_2$Te$_5$, shows a FM order [39]. Theoretical studies predict Mn$_2$(Bi/Sb)$_2$Te$_5$ can host topological surface states [40,41]. Mn$_2$Sb$_2$Te$_5$ is predicted to possess a NiAs-type crystal structure with an AFM ground state which generates gapped Dirac states [40]. With the polarization of magnetic moment by external magnetic field, a topological phase transition into a Weyl state is expected, which is characterized by Weyl point very close to the Fermi level (~ 7 meV) [40]. This is in sharp contrast with the Bi-based counterpart compound Mn$_2$Bi$_2$Te$_5$, which shows a gapped Dirac state in the FM state [40].

Another well-known key feature of Mn-Sb-Te compounds is the intermixing of Mn and Sb atomic sites. This site mixing causes Mn atoms to exist in multiple valence states, such as $Mn^{2+}$ and $Mn^{3+}$ [33,42]. The coexistence of magnetic atoms in different valence states leads to the emergence of complex magnetic phenomena, including multiple magnetic orders, spin glass behaviour, and other frustrated magnetic states [33,43]. Such atomic site mixing can also have a significant impact on the electronic structure, potentially inducing charge carrier localization, similar to the behaviour observed in the $Fe_{1+y}(Te_{1-x}Se_x)$ system [44,45]. However, the influence of site mixing on the magnetic and electrical transport properties of $Mn_2Sb_2Te_5$ remains unexplored to the best of our knowledge. Despite several exotic properties predicted in theoretical calculations, the experimental studies on $Mn_2Sb_2Te_5$ are very few. In this work, we systematically studied the magnetic properties and their correlation with electronic properties of $Mn_2Sb_2Te_5$. AC susceptibility confirmed a spin-glass state, with strong FM correlations. A strong anomalous Hall effect is observed without any topological feature, indicating the absence of ferromagnetism-induced Weyl semi-metallic state. This study provides crucial insights into the complex interplay between magnetism and electronic properties in $Mn_2Sb_2Te_5$, highlighting the spin-glass behaviour and strong FM correlations in shaping its transport properties. This study not only advances the understanding of intrinsic magnetic topological materials but also offers a foundation for exploring novel quantum phases in the Mn-Sb-Te system.

**Experiment**

Single crystals of $Mn_2Sb_2Te_5$ were grown using a self-flux method following Ref. [39]. Mn, Sb, and Te powders were combined in stoichiometric ratios and thoroughly mixed with an agate mortar and pestle. The homogeneous mixture was then pelletized into rectangular shapes, vacuum-sealed in a quartz ampoule, and heated to 1000 ºC over 8 hours, maintained at this temperature for 48 hours, and then slowly cooled to 600 ºC over 120 hours before quenching in water. Magnetization measurements were carried out using a Magnetic Property Measurement System (MPMS3 SQUID, Quantum Design), which enables high-sensitivity magnetization measurements over a wide temperature (400 K – 1.8 K) and magnetic-field range (±7 T). Hall resistivity and specific heat measurements were performed using a Physical Property Measurement System (PPMS, Quantum Design), allowing controlled measurements of transport and thermal properties under variable temperature (400 K – 1.8 K) and magnetic field conditions (±9 T).

**Results and Discussion**

Structural characterization of the $Mn_2Sb_2Te_5$ single crystal was carried out using Rietveld refinement of the powder XRD pattern, as presented in Fig. 1. $Mn_2Sb_2Te_5$ crystallizes in a trigonal structure with *P -3 m 1* space group symmetry, as shown in the inset of Fig. 1. The experimental PXRD pattern shows an excellent match with the calculated profile, yielding goodness-of-fit parameters $R_p$ = 1.56 and $R_{wp}$ = 2.55, both within the acceptable range. The refined lattice parameters are $a = b$ = 4.253(6) Å and $c$ = 17.865(4) Å. $Mn_2Sb_2Te_5$ was reported to have antisite disorder of Sb and Mn atoms via XPS measurement in our earlier report [39], so antisite disorder was taken into account while performing structural refinement. Mn atoms were allowed to partially occupy the Sb (*2d*) site, while Sb atoms were permitted to occupy the Mn (*2c*) site. The refinement confirms substantial antisite disorder at both sites, with approximately 22% Sb atoms occupying the Mn (*2c*) site and 15% Mn atoms occupying the Sb (*2d*) site. The detailed structural parameters are summarized in Table 1.

Figures 2(a) and (b) show the temperature dependence of the magnetization of $Mn_2Sb_2Te_5$ measured under DC magnetic fields along the in-plane (*H*‖*ab*) and the out-of-plane (*H*‖*c*) directions. The magnetic moment begins to increase below 30 K, indicating magnetic ordering of the system. The slightly higher magnetization for *H*‖*c* suggests an easy axis along the c-axis. A peak-like feature is observed around 18 K when $H\|c$, as indicated by the red arrow in Fig. 2b. In our earlier report on $Mn_2Sb_2Te_5$, this anomaly near ~18 K was tentatively interpreted as Neel temperature [39], which was primarily based on DC magnetization measurements and analogy with related Mn-(Bi/Sb)-Te compounds. However, the present measurements suggest that the magnetic behaviour is more complex. For both field orientations, magnetizations measured with field below 1000 Oe exhibit irreversibility between zero field-cooling (ZFC) and field-cooling (FC) data below 15 K, which should be attributed to a spin glass state as will be shown below [33,46,47]. At higher fields above 2500 Oe, the FC and ZFC magnetization converge, both exhibit strong increase up on cooling, indicating strong FM correlations, which is further supported by hysteresis loop in field-dependent magnetization shown below. The observed results are consistent with the theoretical results [38], which indicate a magnetic field-driven FM state in $Mn_2Sb_2Te_5$.

The isothermal magnetization (*M-H*) for $H\|ab$ and $H\|c$ are shown in Figs. 2(c) and (d). For $H\|ab$, magnetization is linear at low field and temperatures, followed by a saturation behaviour at higher field. For $H\|c$, a small hysteresis loop is observed (Fig. 2d, inset) up to 20 K, which, together with the magnetization above 0.6 T, implying strong FM correlations. Above 20 K, magnetization displays linear field dependence, implying a paramagnetic state. Such observation indicates the peak at 18 K in temperature dependent magnetization (red arrow in Fig. 1b) may correspond to magnetic phase transition. Additionally, below 10 K, isothermal magnetization shows a two-step behaviour at 0.5 kOe and 5 kOe, marked by arrows in the inset of Fig. 2(d), which suggests the presence of two magnetic states polarizing at different fields. A similar feature was observed in $Mn_2Bi_2Te_5$ at higher fields [38].

The formation of a spin glass state below 15 K can be demonstrated by the frequency-dependent freezing temperature ($T_f$) in AC susceptibility measurements [48–50]. AC susceptibility measurements were performed with magnetic field along the *c*-axis, using a zero

DC background and an AC field of 5 Oe with frequency $f$ varying from 250 Hz to 10 kHz. Figure 3(a) depicts the variation of real ($\chi'$) and imaginary ($\chi''$) parts of AC susceptibility with temperature. Two prominent peaks at $T_f \sim 11$ K and $T_c \sim 18$ K are observed, corresponding to the freezing temperature and magnetic ordering temperature. Minor features such as a weak shoulder around ~15 K and a broad feature near ~40 K are also visible in the AC susceptibility data. However, these features are not observed in DC magnetization. Therefore, they are unlikely to correspond to intrinsic magnetic phase transitions and are most likely related to background or instrumental effects inherent to AC susceptibility measurements. $T_f$ in both $\chi'$ and $\chi''$ shifts towards higher temperature as the frequency of the AC field is increased - a typical spin glass characteristic. Such a shift with frequency is clearly visible in the inset of Fig. 3(a), which shows the zoomed plot of $\chi'(T)$ near $T_f$. On the other hand, no shift in $T_c \sim 18$ K is observed, implying a possible static nature of the corresponding magnetic phase transition. Similar two peaks in AC susceptibility have been observed for $Mn_{1+x}Sb_{2-x}Te_4$, which were designated to two different spin glass states [42], but in the present case, it is hard to relate both susceptibility peaks to spin glass due to the lack of frequency dependence at $T_c$. In fact, the frequency-independent AC susceptibility peak at $T_c$ followed by the frequency-dependent peak at $T_f$ is reminiscent of reentrant spin-glass systems [51], where the ferromagnetic ordering established below $T_c$ becomes unstable at lower temperatures as randomness and competing exchange interactions dominate. Consequently, the spins gradually lose coherence and freeze into a disordered, non-equilibrium configuration below $T_f$. This transition reflects a crossover from an ordered ferromagnetic to a frozen spin-glass-like state [51]. The observed behaviour is consistent with that reported for well-established systems exhibiting a reentrant spin-glass state, such as magnetic alloys [47,52] and manganites [53]. The observed spin glass feature may stem from the presence of $Mn^{2+}$ and $Mn^{3+}$ mixed ions, as revealed by X-ray photoelectron spectroscopy measurements [39]. Cation mixing, common in the Mn-Sb-Te family [33,42], results in antisite defects where some Mn atoms occupy $Sb^{3+}$ sites, which gives rise to $Mn^{3+}$. The presence of antisite defects in $Mn_2Sb_2Te_5$ is evidenced from the Rietveld refined XRD pattern, which shows that both Mn and Sb atoms are interchanging their atomic positions. Cationic mixing may also induce atomic short-range order, influencing local magnetic interactions. The coexistence of multiple valence states in magnetic ions often leads to spin glass states due to competing magnetic interactions as observed in $MnSb_2Te_4$ [33], manganites [54] and double perovskite [55].

For spin glass, the relative shift in freezing temperature per frequency decade is calculated as $\delta T_f = \frac{\Delta T_f}{T_f \Delta(log_{10}v)}$ where $v = 2\pi f$ is the angular frequency. The obtained $\delta T_f$ of 0.02 lies within the range of $0.0045 \leq \delta T_f \leq 0.08$ for canonical spin glass systems [56]. Further analysis of the spin glass behaviour was conducted by examining the frequency dependence of the freezing temperature $T_f$. Conventionally, this dependence follows a power law [46,56]:

$$\tau = \tau_0 \left(\frac{T_f - T_{SG}}{T_{SG}}\right)^{-n} \tag{1}$$

where $\tau$ is the relaxation time corresponding to the AC frequency ($\tau = 1/2\pi f = 1/v$), $\tau_0$ is the characteristic relaxation time of a single spin flip, $T_{SG}$ is the zero-frequency spin-glass temperature, and the term $\left(\frac{T_f - T_{SG}}{T_{SG}}\right)$ represents reduced temperature $t$. Here $T_{SG} \sim 10.62K$ can be extracted by extrapolating the frequency dependence for $T_f$. Figure 3(b) depicts the experimentally determined relaxation time $\tau$ as a function of the reduced temperature $t$, the fitting of which with the above Eq. 1 yields $\tau_0 = 5.01(6) \times 10^{-8}$ s. This value is much higher than $10^{-12}$ s, the typical value for a canonical spin glass system [49], suggesting the presence of strongly interacting clusters in $Mn_2Sb_2Te_5$. The presence of clusters can be verified by the deviation from the Arrhenius formalism $v = v_0 exp\left(\frac{-E_a}{k_B T_f}\right)$, where $E_a$ denotes the activation energy, and $k_B$ is the Boltzmann constant. This is because that the spin dynamics controlled by single spin flips tends to follow the Arrhenius law [46,56]. However, as shown in Fig. 3(c), the linear dependence between ln($\nu$) and $1/T_f$ expected from the above Arrhenius formalism does not hold for $Mn_2Sb_2Te_5$, especially at the low frequency regime. This deviation implies collective spin freezing in $Mn_2Sb_2Te_5$, which is better modelled by the empirical Vogel-Fulcher law [46,57]:

$$\nu = \nu_0 \exp\left(\frac{-E_a}{k_B(T_f - T_0)}\right) \quad (2)$$

where $T_0$ is the Vogel-Fulcher temperature, and $\nu_0$ is the characteristic frequency that can be extracted from the characteristic relaxation time $\tau_0$ obtained above through $\nu_0 = 1/\tau_0$. As shown in Fig. 3(d), Vogel-Fulcher law reproduces the data very well, yielding an activation energy $E_a$ of 0.97 meV and Vogel-Fulcher temperature $T_0$ of 9.65 K. The non-zero value of $T_0$ indicates the presence of interactions among spins of different magnetic ions, forming a cluster spin glass [49,50,58].

To gain more insight into magnetism in $Mn_2Sb_2Te_5$, heat capacity measurements have been performed to detect bulk magnetic ordering. As shown in Fig. 4, no clear anomaly has been observed in heat capacity near 18 K, where ferromagnetic ordering appears according to magnetic measurements. However, a small deviation between the zero-field and 9 T heat capacity emerges around this temperature (lower inset of Fig. 4) that could be attributed to the onset of ferromagnetic correlations. The absence of a sharp $\lambda$-type anomaly suggests that the magnetic transition is broad and involves a gradual development of magnetic order rather than a well-defined long-range transition. Such behaviour is characteristic of reentrant cluster spin-glass systems [53,59], where the total magnetic contribution to the specific heat is small and indicates a small change in entropy at the transition.

Ignoring magnetic contributions, the low-temperature heat capacity $C$ can be expressed as

$$C = \gamma T + \beta T^3 \quad (3)$$

where $\gamma$ is the Sommerfeld coefficient and related to the electronic contribution $\gamma T$ to heat capacity, and $\beta T^3$ is the phonon contribution. Providing that the magnetic contribution is rather weak as demonstrated by the negligible impact of magnetic field on the heat capacity, the above Eq. 3 was used to analysis the measured heat capacity. As shown in the upper inset of Fig. 3, the fit of $C/T$ to $\gamma + \beta T^2$ yield $\gamma = 0.14(1)$ Jmol$^{-1}$K$^{-2}$ and $\beta = 0.0036(3)$ Jmol$^{-1}$K$^{-2}$. From the obtained $\beta$, Debye temperature $\theta_D$ for Mn$_2$Sb$_2$Te$_5$ was calculated as 116±1 K via $\theta_D = \left(\frac{12\pi^4 nR}{5\beta_n}\right)^{1/3}$ where $R$ is the ideal gas constant. The obtained value of $\gamma$ is quite high as compared to normal metals. Such enhanced $\gamma$ value is comparable to that of reported spin glass systems [49,60–62], which can be understood in terms of competing magnetic interactions that gives rise to a high degree of frustrations [56].

Mn$_2$Sb$_2$Te$_5$ exhibits various magnetic phases, which have been summarized in the field ($H$)-temperature ($T$) phase diagram shown in Fig. 5. This phase diagram shows the evolution from the paramagnetic state into a regime of short-range ferromagnetic correlations below $T_c \approx 18$ K, followed by reentrant freezing into a ferromagnetic cluster spin-glass state below $T_f \approx 11$ K at low fields. It also illustrates the suppression of glassy freezing and the emergence of a field-polarized FM-like state under an applied magnetic field of 2.5 kOe. The observed magnetic properties in Mn$_2$Sb$_2$Te$_5$ are in sharp contrast with its Bi analogue, which shows a clear AFM transition. This might be attributed to the nature of the present antisite defect. Mn-Sb antisite disorder is expected to be more pronounced than Mn-Bi antisite disorder [63], which enhances Mn/Sb site mixing and mixed Mn$^{2+}$/Mn$^{3+}$ valence, introduces competing magnetic exchange interactions, which frustrate long-range magnetic order and stabilizes a glassy magnetic ground state in $Mn_2Sb_2Te_5$.

With the understanding of magnetism in $Mn_2Sb_2Te_5$, its impact on transport is studied via the anomalous Hall effect. The magnetic field $(\mu_0H)$ dependent transverse (Hall) resistivity $(\rho_{yx})$ is shown in Fig. 6(a). The positive slope indicates *p*-type charge carriers in $Mn_2Sb_2Te_5$, which is in sharp contrast to $Mn_2Bi_2Te_5$, that possess electron-type carriers [38]. A similar change in carrier type has also been observed in $MnBi_2Te_4$ [8] and $MnSb_2Te_4$ [64]. Carrier concentration, based on the Hall resistivity slope, is approximately $3.67\times10^{20}$ cm$^{-3}$ at 2 K. Near zero field, the non-zero Hall resistivity and Hall resistivity loop (Fig. 6b) reflect an anomalous Hall effect (*AHE*). The close resemblance between Hall resistivity loop (Fig. 6b) and the magnetization loop (Fig. 3b), as well as their similar temperature dependence, indicates a magnetic origin for *AHE*. To further elaborate on the coupling between magnetic and electrical properties in the studied $Mn_2Sb_2Te_5$, the derivative of Hall resistivity and magnetization with respect to $\mu_0H$ has been plotted together in the inset of Fig. 6(a). The field-induced steps in Hall resistivity align with magnetization measurements, as shown by arrows in the inset of Fig. 6(a). Similar coupled transport and magnetization properties were also observed in its sister compound, $Mn_2Bi_2Te_5$ [38]. Interestingly, despite the absence of long-range magnetic order, $Mn_2Sb_2Te_5$ exhibits a strong anomalous Hall effect, which confirms the presence of robust FM correlations revealed in magnetization measurements. The observed AHE demonstrate that robust anomalous Hall responses can emerge from glassy and cluster-based magnetic states, and highlights that magnetic disorder and short-range correlations can strongly influence electronic transport in magnetic topological material candidates. Theoretically, the FM $Mn_2Sb_2Te_5$ is predicted to host a Weyl phase [40], which is reminiscent of the FM Weyl state up on the polarization of magnetic moment under magnetic field in $Mn(Bi,Sb)_2Te_4$ [26,65,66]. In $Mn(Bi,Sb)_2Te_4$, the transition to a FM Weyl state is manifested in a strong intrinsic topological Hall effect [26]. However, the signature of topological Hall effect is not observed in $Mn_2Sb_2Te_5$. This might be due to the fact that the Weyl points are not close to the Fermi level, which is supported by rather high hole carrier density of $10^{20}$ cm$^{-3}$. Such a large carrier density implies that the chemical potential is pushed far away from the low-energy band features. As an order-of-magnitude estimate, treating the holes as a three-dimensional degenerate Fermi gas, Fermi energy can be calculated using the relation $E_F = \frac{\hbar^2 k_F^2}{2m^*}$, where $k_F$ is Fermi wavevector and can be calculated by the hole density using the relation $p = \frac{k_F^3}{3\pi^2}$, for this analysis bare electron mass has been considered. For $p = 3.67 \times 10^{20}$ cm$^{-3}$, $E_F$ is found to be ~ 0.17 eV, much larger than the ~0.7 meV associated with Weyl nodes predicted in the ideal ferromagnetic state [40]. Consequently, any Weyl-related Berry-curvature contributions to transport are expected to be strongly masked by the large background of conventional carriers, and the Hall response is therefore dominated by the ordinary and anomalous Hall components, consistent with the experimental observations.

This is in agreement with $Mn(Bi,Sb)_2Te_4$, where clear Weyl-related transport signatures have been reported only when the carrier density is reduced to ~ $10^{18}$ cm$^{-3}$ through careful Bi/Sb composition tuning [26]. Sb-rich compositions, which retain similarly high hole densities, do not show clear evidence of Weyl transport features. Motivated by this established trend, partial substitution of Sb by Bi in $Mn_2(Bi/Sb)_2Te_5$ provides a plausible route to lower the carrier density and shift the chemical potential closer to the Weyl nodes, thereby improving the prospects for realising a magnetic Weyl semimetal state in this material family.

**Conclusion:**

In summary, we have demonstrated the magnetic, thermal, and electronic properties of $Mn_2Sb_2Te_5$, a potential new member to the family of magnetic topological materials. Unlike its Bi counterpart, $Mn_2Bi_2Te_5$ which exhibits AFM ordering, $Mn_2Sb_2Te_5$ displays a more complex magnetism characterized by a spin glass state with strong FM correlations that is distinct from the predicted ground state. This complexity might be attributed to atomic site intermixing between Mn and Sb atoms. Additionally, the transverse magneto-transport measurement reveals AHE, highlighting the coupling between magnetic and electronic properties. However, the Weyl phase, which is anticipated in the FM phase of $Mn_2Sb_2Te_5$, remains elusive, which is likely because the Weyl point is too far from the Fermi energy as revealed by high carrier density. This study emphasizes the crucial role of atomic disorders in shaping the properties of $Mn_2Sb_2Te_5$. Further investigations, such as Bi substitution to lower the carrier density, could help in tuning its electronic and magnetic properties, potentially revealing the anticipated topological states and deepening the understanding of its topological nature.

**Acknowledgement:**

M.M. Sharma acknowledges support by µ-ATOMS, an Energy Frontier Research Center funded by DOE, Office of Science, Basic Energy Sciences, under Award DE-SC0023412. The use of MPMS3 SQUID is supported by MonArk NSF Quantum Foundry, which is supported by the National Science Foundation Q-AMASE-i program under NSF award No. DMR-1906383. Ankush Saxena acknowledges the financial support by the Ministry of Science and Technology R.O.C. Taiwan for Grant No. MOST-113-2112-M-110-018.

**Data Availability Statement:**

The data will be made available from the corresponding author upon reasonable request.

**Table:** Structural parameters of $Mn_2Sb_2Te_5$ obtained from Rietveld refinement of PXRD pattern.

| Compound | Lattice parameters (Å) | | | Atoms | Wycoff | Atomic positions | | | Occ. |
|---|---|---|---|---|---|---|---|---|---|
| | $a$ | $b$ | $c$ | | | $x$ | $y$ | $z$ | |
| $Mn_2Sb_2Te_5$ | 4.253(7) | 4.253(7) | 17.870(0) | Mn1 | $2c$ | 0.0000 | 0.0000 | 0.3463 | 0.775 |
| | | | | Sb1 | $2c$ | 0.0000 | 0.0000 | 0.3463 | 0.220 |
| | | | | Mn2 | $2d$ | 0.3333 | 0.6666 | 0.0823 | 0.150 |
| | | | | Sb2 | $2d$ | 0.3333 | 0.6666 | 0.0823 | 0.854 |
| | | | | Te1 | $1a$ | 0.0000 | 0.0000 | 0.0000 | 1.000 |
| | | | | Te2 | $2d$ | 0.3333 | 0.6666 | 0.2049 | 1.000 |
| | | | | Te3 | $2d$ | 0.3333 | 0.6666 | 0.4274 | 1.000 |

**Figures:**

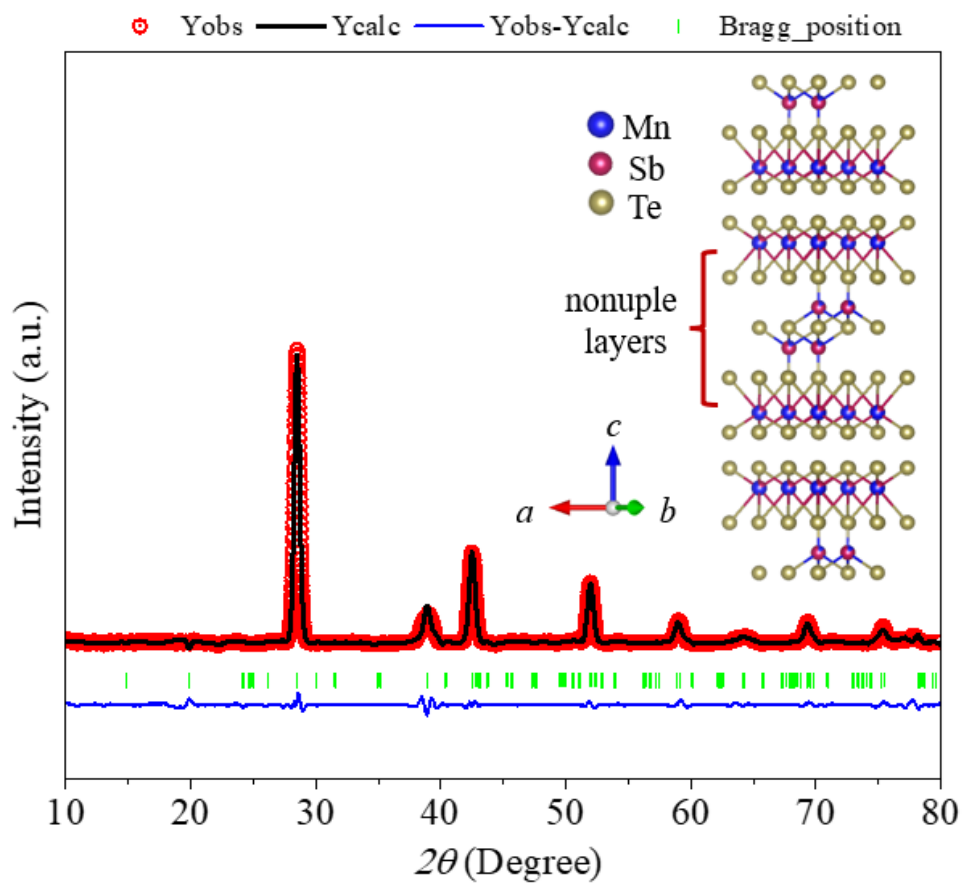

**Fig. 1:** Rietveld refined XRD pattern of Mn$_2$Sb$_2$Te$_5$ fitted with *P -3 m 1* space group symmetry considering antisite disorder on Mn (*2c*) and Sb (*2d*) atomic sites, the inset is showing the corresponding unit cell showing partial occupancies on Mn (*2c*) and Sb (*2d*) atomic sites.

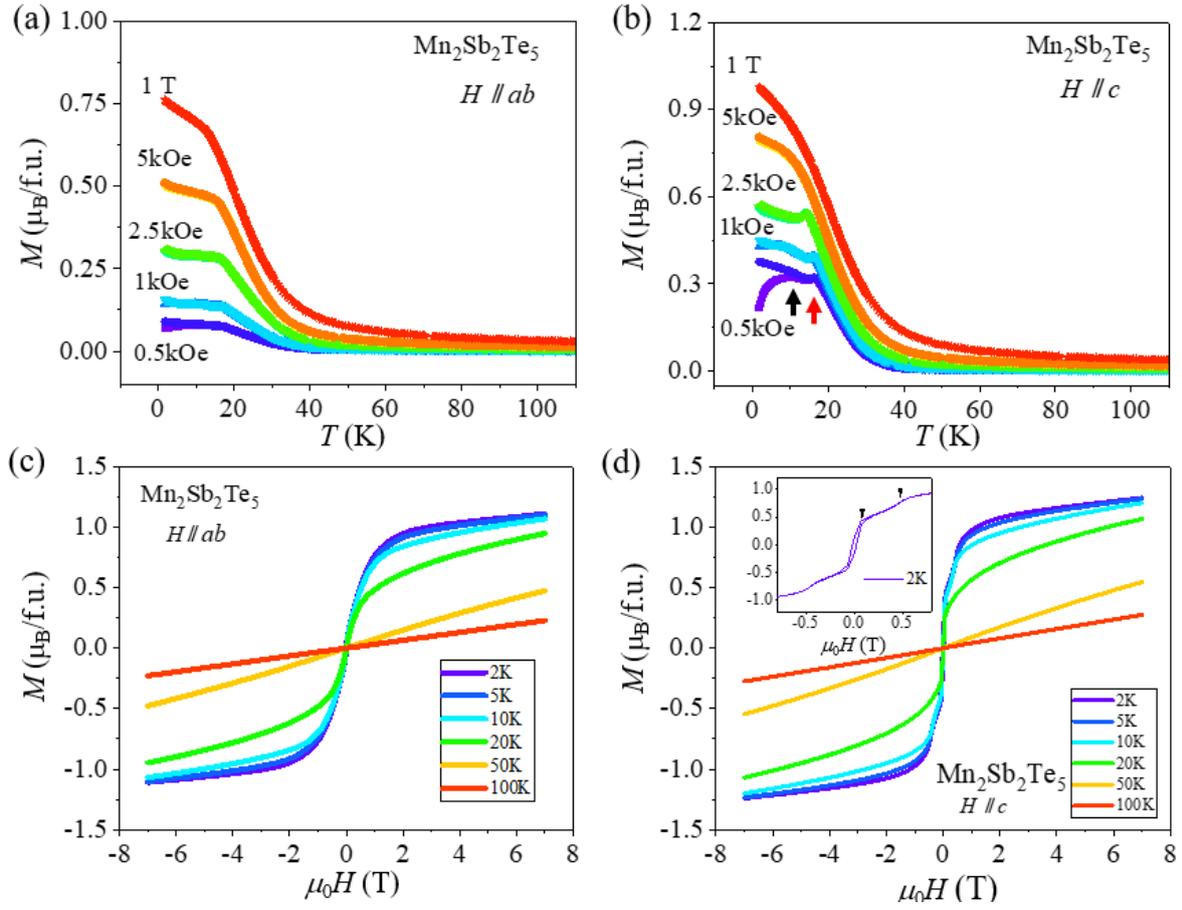

**Fig. 2:** DC magnetic moment (*M*) vs temperature (*T*) curve of Mn$_2$Sb$_2$Te$_5$ under various magnetic fields directed along (a) *ab* plane (b) *c*-axis. Isothermal magnetization vs applied magnetic field ($\mu_0H$) curve of Mn$_2$Sb$_2$Te$_5$ at various temperatures with the applied magnetic field aligned along (c) *ab* plane (d) *c* axis.

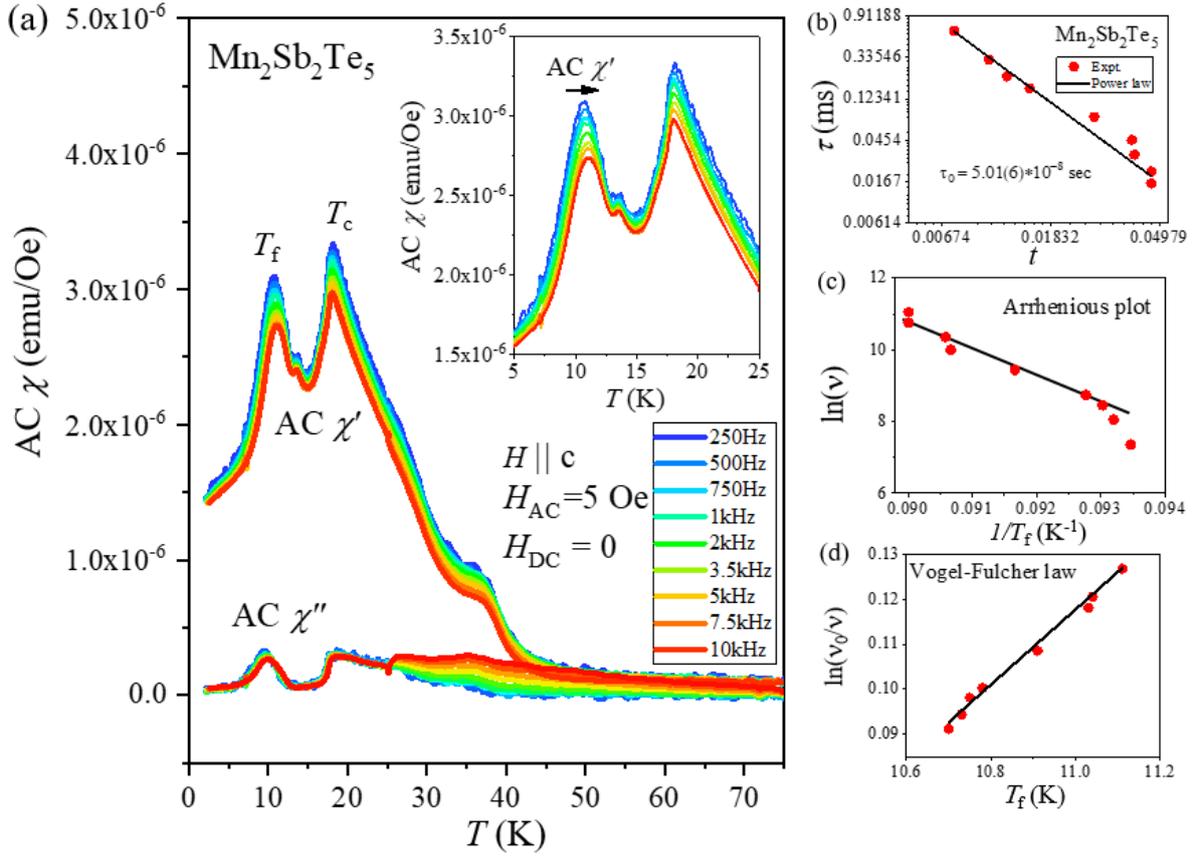

**Fig. 3:** (a) AC susceptibility vs T measurement of $Mn_2Sb_2Te_5$ single crystal performed under zero DC background an AC magnetic field of 5 Oe, while the frequency is changed from 250 Hz to 10 kHz. The inset shows the variation in real part of AC susceptibility (AC $\chi'$) with respect to temperature in the vicinity of spin glass transition at different AC frequencies. (b) The frequency dependence of freezing temperature ($T_f$) plotted as relaxation time ($\tau$) vs reduced temperature ($t$). (c) The frequency dependence of $T_f$ is plotted as $\ln(\nu)$ vs $1/T_f$, which is fitted with the Arrhenius law shown as black solid line. (d) The frequency dependence of $T_f$ plotted as $T_f$ vs $\ln(\nu_0/\nu)$, and fitted with Vogel-Fulcher law shown by black solid line.

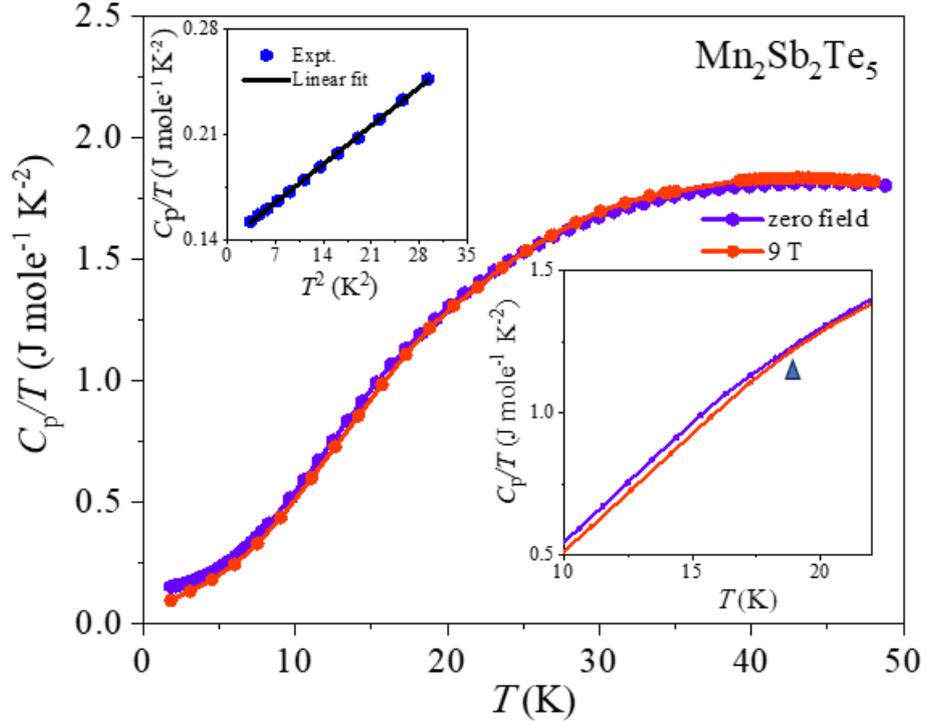

**Fig. 4:** Temperature dependence of specific heat plotted as $C_p/T$ vs $T$, under the magnetic field of 0 T and 9 T. The upper left inset shows the $C_p/T$ vs $T^2$ plot, fitted linearly as shown by the black solid line. The lower right inset shows the zoomed view of $C_p/T$ vs $T$ plot exhibiting the deviation of specific heat under zero magnetic field and a magnetic field of 9 T, evidencing the contribution of magnetism in specific heat of $Mn_2Sb_2Te_5$.

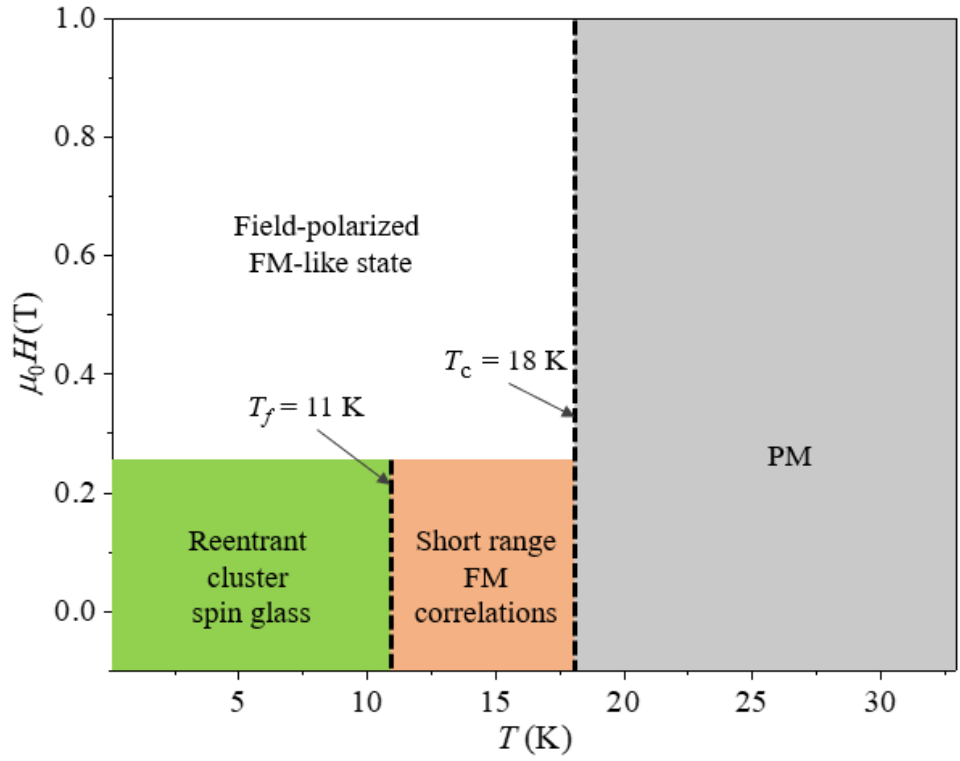

**Fig. 5:** *H-T* phase diagram describing different magnetic phases of $Mn_2Sb_2Te_5$.

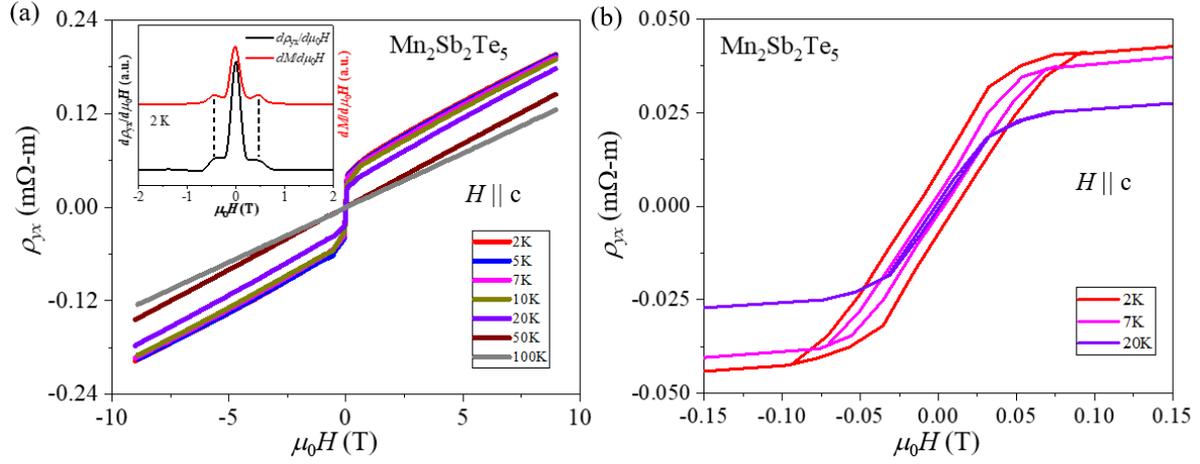

**Fig. 6:** (a) Hall resistivity ($\rho_{yx}$) as a function of applied magnetic field ($\mu_0 H$) at various temperatures, showing the presence of finite Hall resistivity in the absence of magnetic field evidencing the presence of the Anomalous Hall effect. The inset shows the first-order derivative of magnetization and Hall resistivity with respect to the applied magnetic field. (b) Zoomed view of $\rho_{yx}$ vs $\mu_0 H$ plot in the vicinity of zero magnetic field showing the presence of hysteresis emerged due to intrinsic magnetization in $Mn_2Sb_2Te_5$.